\documentstyle[twoside,fleqn,epsfig,espcrc2]{article}

\title{Universal and non-universal behavior in Dirac spectra}

\author{M.E.~Berbenni-Bitsch\address{Fachbereich Physik--Theoretische
    Physik, Universit\"at Kaiserslautern, D-67663 Kaiserslautern,
    Germany}\thanks{Poster presented by M.E.~Berbenni-Bitsch}, 
  M.~G\"ockeler\address{Institut f\"ur Theoretische Physik,
    Universit\"at Regensburg, D-93040 Regensburg, Germany},
  S.~Meyer$^{\rm a}$,
  A.~Sch\"afer$^{\rm b}$, and
  T.~Wettig\address{Institut f\"ur Theoretische Physik, Technische
    Universit\"at M\"unchen, D-85747 Garching,
    Germany}\thanks{Talks presented by T.~Wettig}}

\begin{document}

\begin{abstract}
  We have computed ensembles of complete spectra of the staggered
  Dirac operator using four-dimensional SU(2) gauge fields, both in
  the quenched approximation and with dynamical fermions.  To identify
  universal features in the Dirac spectrum, we compare the lattice
  data with predictions from chiral random matrix theory for the
  distribution of the low-lying eigenvalues.  Good agreement is found
  up to some limiting energy, the so-called Thouless energy, above
  which random matrix theory no longer applies.  We determine the
  dependence of the Thouless energy on the simulation parameters using
  the scalar susceptibility and the number variance.
\end{abstract}

\maketitle

\section{Introduction}
\label{intro}

The low-energy particle spectrum implies that the chiral symmetry
which the QCD Lagrangian possesses in the limit of massless quarks is
spontaneously broken.  The corresponding order parameter, the chiral
condensate, can be related to the pion decay constant through the
Gell-Mann--Oakes--Renner relation \cite{Gell68}.  On the other hand,
the chiral condensate is directly related to the density of the
smallest eigenvalues of the Dirac operator via the Banks-Casher
relation \cite{Bank80}.

On the lattice, the density of the low-lying eigenvalues of the Dirac
operator is strongly dependent on the choice of the fermion action.
The staggered fermion action respects a chiral ${\rm U}_{\rm A}(1)$
symmetry and, as a consequence, the nonzero eigenvalues of the Dirac
operator occur in pairs $\pm \lambda_{n}$.  Therefore, the eigenvalues
near zero and in the bulk of the spectrum can be distinguished.  The
Wilson fermion action breaks chiral symmetry explicitly, and there is
no ${\rm U}_{\rm A}(1)$ symmetry for the hermitian Wilson Dirac
operator.

In this contribution, we will be concerned with universal features in
the spectrum of the staggered Dirac operator.  We will mainly
concentrate on the low-lying eigenvalues since they are particularly
relevant for chiral symmetry breaking.  The Banks-Casher relation
\cite{Bank80}, $\pi\rho(0)=V\Sigma$, relates the eigenvalue density,
$\rho(\lambda)=\langle\sum_n\delta(\lambda-\lambda_n)\rangle$, of the
Dirac operator at zero virtuality to the absolute value of the chiral
condensate, $\Sigma$.  $V$ is the four-volume.  If chiral symmetry is
spontaneously broken, this relation implies that the spacing of the
small eigenvalues is $\sim1/(V\Sigma)$.  Since the eigenvalues of the
non-interacting Dirac operator are spaced like $1/V^{1/4}$, this means
that the Dirac eigenvalues in QCD must be strongly correlated.

Leutwyler and Smilga \cite{Leut92} derived an effective low-energy
theory which is valid in the range $1/\Lambda<L<1/m_\pi$, where $L$ is
the linear extent of the box, $\Lambda$ is a typical hadronic scale,
and $m_\pi$ is the pion mass. In this region, the kinetic terms in the
chiral Lagrangian can be neglected, and only the symmetries are
important.  It was then conjectured by Shuryak and Verbaarschot
\cite{Shur93} that the distribution of the small Dirac eigenvalues,
averaged over gauge field configurations, is universal in the sense
that it only depends on global symmetries of the Dirac operator.  The
essential ingredient is the spontaneous breaking of chiral symmetry in
the QCD vacuum.  Given this fact, the statistical properties of the
low-lying eigenvalues can be computed in a much simpler theory, e.g.,
chiral random matrix theory (RMT).  Alternatively, one can use the
finite-volume partition function of Leutwyler and Smilga.  We will not
address this point since it is discussed in the contribution by P.H.
Damgaard \cite{Damg98}.

To resolve the low-lying Dirac eigenvalues, one rescales the energies
by a factor of $V\Sigma$ and defines the microscopic spectral density
\cite{Shur93},
\begin{equation}
  \label{eq1.0}
  \rho_s(z)=\lim_{V\rightarrow \infty}\frac{1}{V\Sigma}
  \rho\left(\frac{z}{V\Sigma}\right)\:.
\end{equation}
This is a typical universal quantity which can be computed in chiral
RMT.  Other universal quantities are the distribution of the smallest
eigenvalue and higher order spectral correlation functions on the
microscopic scale $\lambda\sim1/(V\Sigma)$.

We note that by concentrating on the low-lying eigenvalues, we are
considering a finite-volume effect.  In particular, by computing
analytical results we hope to learn something about the approach to
the thermodynamic limit.  On the lattice, one necessarily has to
perform an extrapolation to this limit.  An example of the utility of
the RMT results with regard to the thermodynamic limit has already
been given in Ref.~\cite{Berb98b}.

We briefly review chiral random matrix theory in Sec.~\ref{chRMT} and
describe numerical details of our simulations in Sec.~\ref{numerics}.
Lattice data for the small Dirac eigenvalues obtained in the quenched
approximation are compared with RMT predictions in Sec.~\ref{quen}.
New results with dynamical fermions are discussed in Sec.~\ref{dyna}.
These are particularly interesting with regard to the chiral limit.
Of course, the random-matrix approach only works in a limited energy
range.  In Sec.~\ref{susc}, we will identify the domain of validity
quantitatively.  We conclude with a summary and an outlook to future
work in Sec.~\ref{summary}.

\section{Chiral random matrix theory}
\label{chRMT}

The Dirac operator in the continuum is defined as
$D=\gamma_\mu(\partial_\mu+gA_\mu)$, where $A$ denotes the gauge field
and $g$ is the coupling constant.  This operator is anti-hermitian.
Because of the ${\rm U}_{\rm A}(1)$ symmetry $\{\gamma_5,D\}=0$, all
nonzero eigenvalues of $iD$ come in pairs $\pm\lambda_n$.  In a chiral
basis, the Dirac matrix has the structure
\begin{equation}
  \label{eq1.1}
  iD=\left[\matrix{0&T\cr T^\dagger&0}\right]\:,
\end{equation}
where $T$ is some complicated matrix.  The Euclidean QCD partition
function in a sector of topological charge $\nu$ is given by
\begin{eqnarray}
  \label{eq1.2}
  \lefteqn{Z_{\rm QCD}^{(\nu)}=
  \int{\cal D}A^{(\nu)}e^{-S_{\rm gl}}
  \prod_{f=1}^{N_f} \det(D+m_f)}\nonumber\\
  &&=\int{\cal D}A^{(\nu)}e^{-S_{\rm gl}}
  \prod_{f=1}^{N_f}m_f^{|\nu|}\!\!
  \prod_{\lambda_n>0}(\lambda_n^2+m_f^2)\,,
\end{eqnarray}
where $S_{\rm gl}$ is the gluonic action, $N_f$ is the number of quark
flavors with masses $m_f$, and the $\lambda_n$ are the eigenvalues of
$iD$.  The superscript $(\nu)$ on $A$ means that the path integral is
only over gauge fields with fixed topological charge $\nu$.  The
partition function is then given by $Z_{\rm QCD}=\sum_\nu
e^{i\theta\nu}Z_{\rm QCD}^{(\nu)}$, where $\theta$ is the vacuum
angle.

The basic idea of the random matrix approach is to substitute the
matrix $T$ in Eq.~(\ref{eq1.1}) by a random matrix $W$, respecting the
global symmetries of the problem.  If the dimension of the matrix $W$
is $N\times(N+\nu)$, there are $\nu$ exact zero modes.  The RMT
partition function reads
\begin{equation}
  \label{eq1.3}
  Z_{\rm RMT}^{(\nu)}=\int{\cal D}WP_0(W)\prod_{f=1}^{N_f}
  \det(WW^\dagger+m_f^2)\:,
\end{equation}
where the precise form of the distribution, $P_0(W)$, replacing
$e^{-S_{\rm gl}}$ is unimportant, provided that it is invariant under
rotations of $W$ and that we are interested in universal properties
\cite{Akem97}.  For convenience, one often uses a Gaussian
distribution.  Eventually, one is interested in the limit $N\to\infty$
which can be identified with the thermodynamic limit.  The
classification of the random matrix ensembles pertaining to different
symmetry classes can be found in Ref.~\cite{Verb94a}.  We will discuss
staggered fermions with gauge group SU(2) which are described by the
chiral symplectic ensemble (chSE) of RMT.  In this ensemble, the
elements of $W$ are real quaternions.  Diagonalizing $W$ and
expressing it in terms of angles and radial coordinates, the
distribution of $W$ can be written as \cite{Verb94a}
\begin{eqnarray}
  \label{eq1.4}
  \lefteqn{P(W)=
  P_0(W)\prod_{f=1}^{N_f}\det(WW^\dagger+m_f^2)}\nonumber\\
  &&\!\!\!\!\!\!\!=P_0(\{\lambda\})\,\Delta^4(\lambda^2)
  \prod_n\lambda_n^{4|\nu|+3}\prod_{f=1}^{N_f}(\lambda_n^2+m_f^2)\:,
\end{eqnarray}
where $\Delta(x)=\prod_{i>j}(x_i-x_j)$ is the Vandermonde determinant.
The main mathematical problem then consists in performing integrations
of this expression over all but a few variables $\lambda_n$ in the
limit $N\to\infty$.

\begin{figure*}
  \centerline{\epsfig{figure=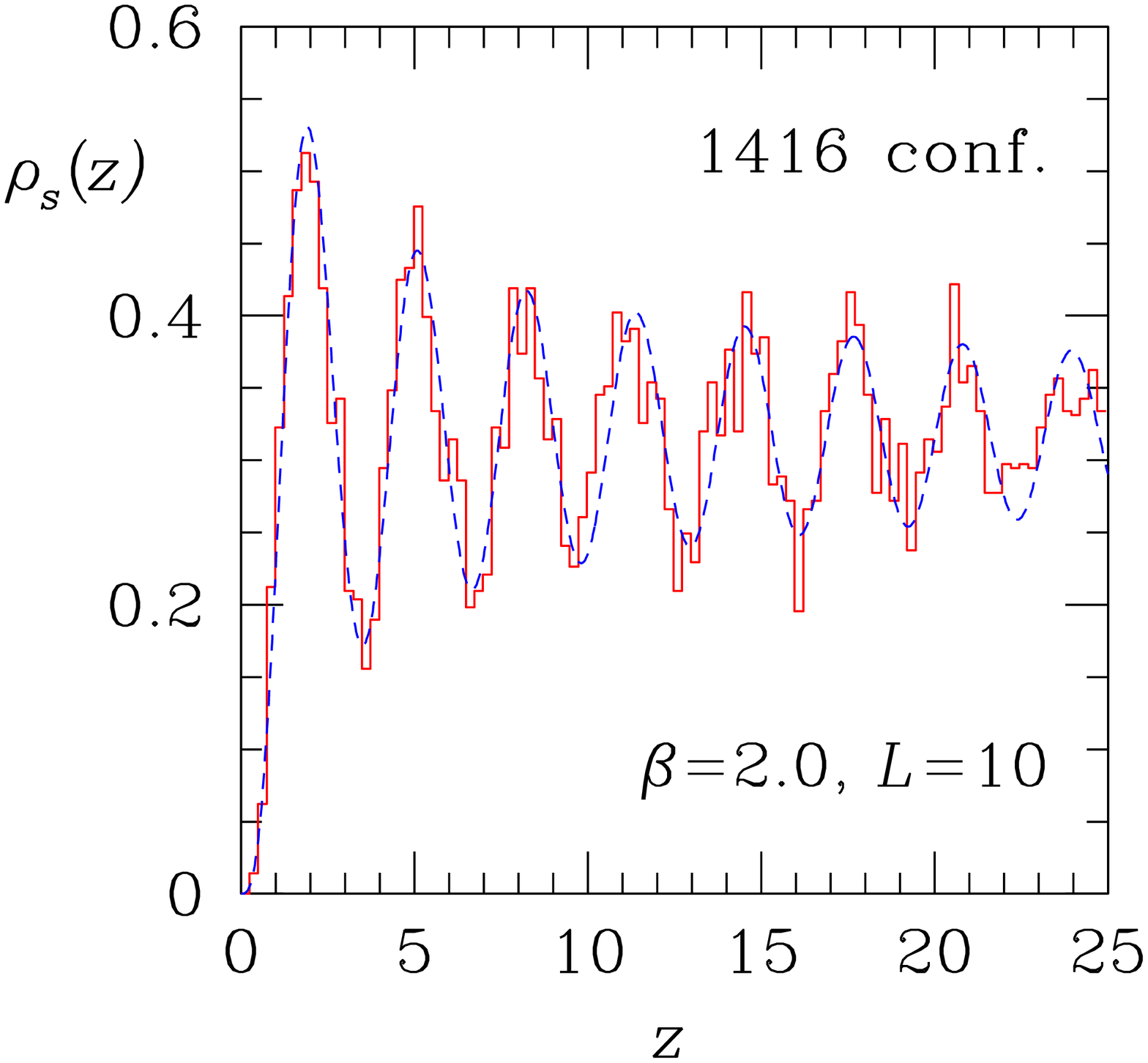,width=50mm}\hspace*{3mm}
    \epsfig{figure=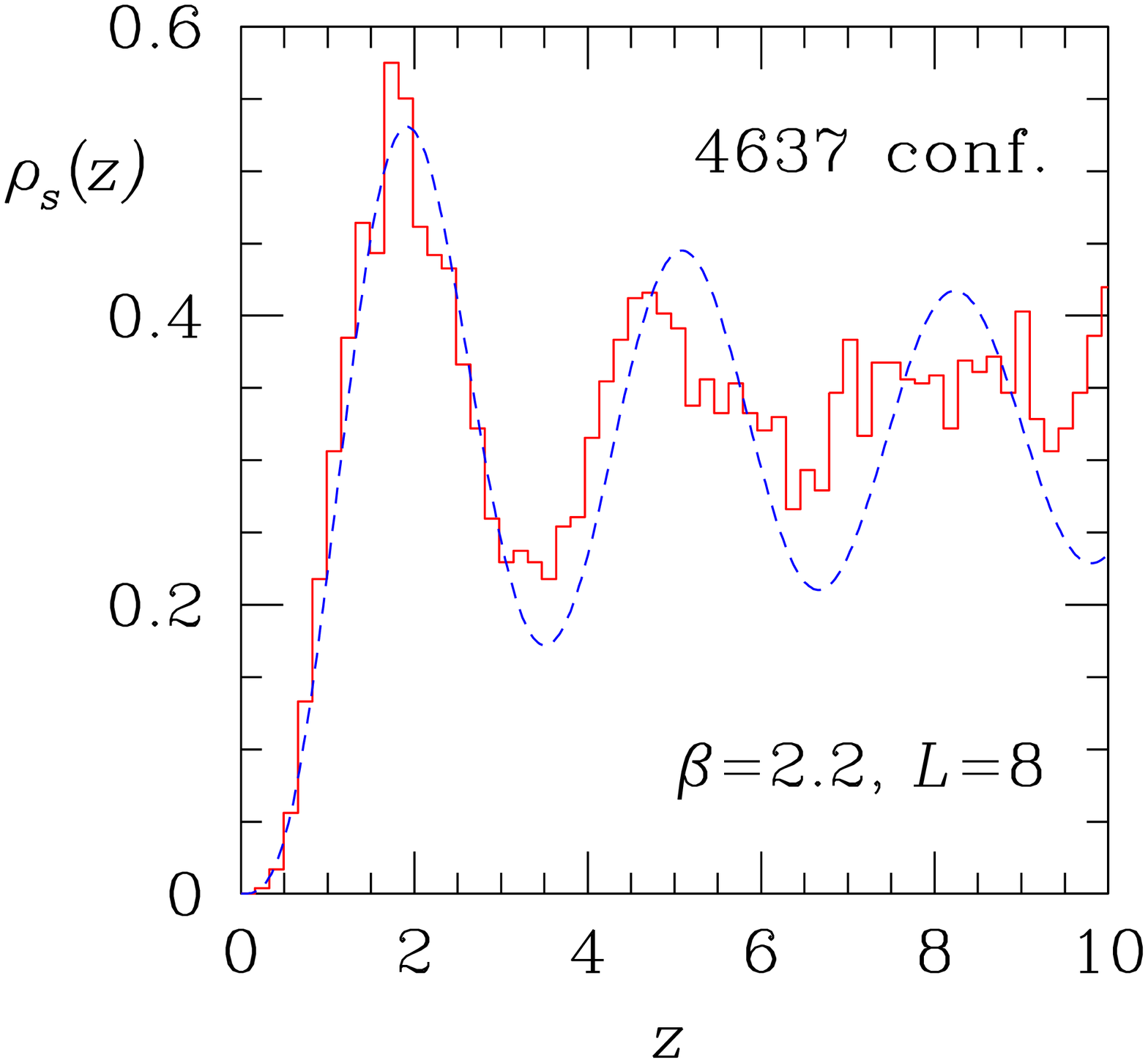,width=50mm}\hspace*{3mm}
    \epsfig{figure=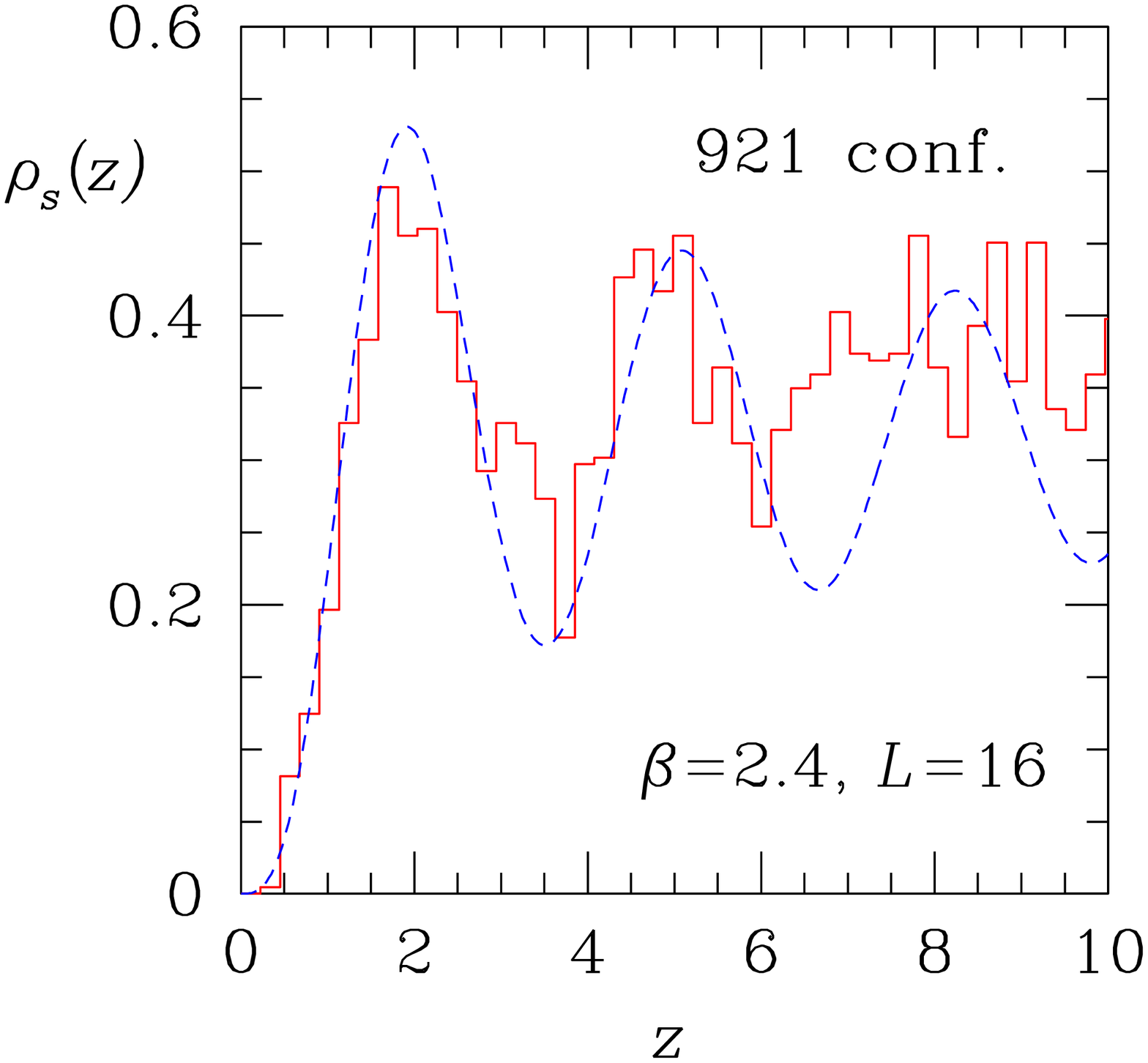,width=50mm}} \vspace*{-8mm}
  \caption{Microscopic spectral density of the lattice Dirac operator
    (histograms) and RMT prediction (dashed lines) for three different
    parameter sets.  Lattice size $L^4$, inverse coupling constant
    $\beta=4/g^2$, and number of configurations are shown in the
    figures.}
  \label{fig1}
\end{figure*}

\section{Numerical details}
\label{numerics}

To check RMT predictions, one often replaces an ensemble average by a
spectral average which is possible due to spectral ergodicity.  Here,
however, we are interested in the low-lying eigenvalues, and a
spectral average is not possible.  Thus, we need good statistics.

Recently, it has become feasible to calculate large ensembles of
complete spectra of the staggered Dirac operator in SU(2)
\cite{Berb98a} on lattices of size up to $16^4$, extending the methods
developed by Kalkreuter \cite{Kalk95}.  We determine the complete
spectrum in order to make sure that the distribution of the low-lying
eigenvalues has no other numerical uncertainties than the finite
precision in 64 bit arithmetic.

Our numerical simulations were done on a CRAY T3E.  The configurations
were generated using a hybrid Monte Carlo algorithm. The boundary
conditions were periodic for the gauge fields and periodic in space
and anti-periodic in Euclidean time for the fermions.  For the
diagonalization of the Dirac matrix, we employed the Cullum-Willoughby
version of the Lanczos algorithm \cite{Kalk95}. In SU(2) with
staggered fermions, every eigenvalue of $iD$ is twofold degenerate
because of a global charge conjugation symmetry. In addition, the
squared Dirac operator $-D^2$ couples only even to even and odd to odd
lattice sites, respectively.  Thus, on a lattice with $N$ sites,
$-D^2$ has $N/2$ distinct eigenvalues.  There is an exact sum rule for
the distinct eigenvalues of $-D^2$,
\begin{equation}
  \label{sumrule}
  \sum_{\lambda_n>0}(\lambda_na)^2=N\:,
\end{equation}
where $a$ is the lattice constant.  Since we generated complete
spectra, we could use this sum rule to check the accuracy of our
eigenvalues.  Equation~(\ref{sumrule}) was satisfied with a relative
precision of about $10^{-8}$.  In Table~\ref{table1}, we summarize our
quenched spectra.  As an example, for $L=16$ the eigenvalue spectrum
has 32,768 different elements spanning four orders of magnitude.  From
Fig.~\ref{fig2} below one can read off that the fluctuations of the
smallest eigenvalue cover one order of magnitude.
\begin{table}
  \begin{center}
    \begin{tabular}{rrcll}
      \hline \\[-3mm]
      \multicolumn{1}{c}{$\beta$}&\multicolumn{1}{c}{$L$}&\# of conf.&
      \multicolumn{1}{c}{$\langle\lambda_{\rm min}\rangle$}&
      \multicolumn{1}{c}{$\tau_{\rm int}$} \\[1mm] \hline \\[-3mm]
      1.8 &  8 & 1999           & 0.00295(3) & 0.69(7) \\
      2.0 &  4 & 9979           & 0.0699(5)  & 1.3(1)  \\
      2.0 &  6 & 4981           & 0.0127(1)  & 0.69(5) \\
      2.0 &  8 & 3896           & 0.00401(3) & 0.71(6) \\
      2.0 & 10 & 1416           & 0.00164(2) &  0.7(1) \\
      2.2 &  6 & 5542           & 0.0293(3)  &  1.7(2) \\
      2.2 &  8 & 2979           & 0.0089(1)  &  1.2(2) \\
      2.4 & 16 & \phantom{0}921 & 0.00390(9) & 1.2(3)  \\
      2.5 &  8 & \phantom{0}576 & 0.194(9)   & 8(3)    \\
      2.5 & 16 & \phantom{0}543 & 0.016(2)   & 12(7)   \\[1mm] \hline
    \end{tabular}
  \end{center}
  \caption{Summary of our quenched spectra with $\beta=4/g^2$ and
      $V=L^4$.  The last two columns represent the average value of
    the smallest eigenvalue (in units of $1/2a$) and its integrated
    autocorrelation time, respectively.} 
  \label{table1}
\end{table}

\section{Quenched results}
\label{quen}

Our first studies were done in the quenched approximation where the
fermion determinants in Eq.~(\ref{eq1.4}) are absent.  Most of the
results were reported in Refs.~\cite{Berb98a,Berb98b,Ma98}.  In
Fig.~\ref{fig1}, we show the microscopic spectral density obtained for
three different sets of parameters, covering strong and weak coupling.
To compare the data with the RMT prediction [Eq.~(\ref{eq4.1}) below
with $\alpha=0$] one needs to compute the parameter $V\Sigma$ which
sets the energy scale.  This parameter is determined by the data via
the Banks-Casher relation, $V\Sigma=\pi\rho(0)$.  [Since we have the
complete spectrum for many configurations, we can easily fit
$\rho(0)$].  Thus, the comparison of Fig.~\ref{fig1} is
parameter-free.

The microscopic spectral density has an oscillatory structure.  The
maxima correspond roughly to the most likely positions of the
individual eigenvalues.  One can also compute the distribution of the
smallest eigenvalue alone, or higher order spectral correlation
functions.  For details, we refer to Refs.~\cite{Berb98a,Ma98}.

\begin{figure}[!b]
  \centerline{\epsfig{figure=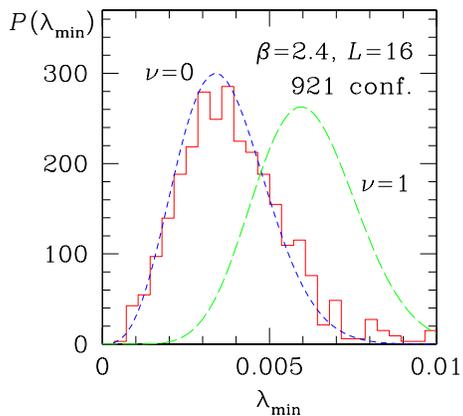,width=60mm}}
  \caption{Distribution of the smallest eigenvalue of the lattice
    Dirac operator (histogram) and RMT predictions for topological
    sectors $\nu=0$ (short dashes) and $\nu=1$ (long dashes) for
    $\beta=2.4$ on a $16^4$ lattice.}
  \label{fig2}
\end{figure}

The data agree with the RMT prediction for topological charge $\nu=0$
(an example is shown in Fig.~\ref{fig2}).
This is due to the fact that in the derivation of the RMT results for
$\nu\ne0$, it is assumed that the Dirac operator has $|\nu|$ exact
zero modes.  On the lattice, the would-be zero modes (in lattice
units) are shifted by an amount proportional to $a^2$.  Therefore,
unless $a$ is very small or some form of improvement is used, we
expect to find agreement of the lattice data with the RMT results for
$\nu=0$.

We can see from Fig.~\ref{fig1} that the agreement of the data with
the RMT result breaks down at some value of the rescaled energy.  The
domain of validity of RMT depends on the lattice volume and on the
coupling constant.  A quantitative discussion of this issue will be
given in Sec.~\ref{susc}.  From the figures, one can see that the
range of validity increases with increasing lattice volume and
decreasing $\beta$.

\section{Results with dynamical fermions}
\label{dyna}

The results obtained in the quenched approximation are very
encouraging.  The natural next step is to include the fermion
determinants in Eq.~(\ref{eq1.4}).  Now an additional parameter enters
the problem, the quark mass (assuming that we consider degenerate
quarks).  Again, we are interested in the small eigenvalues of
magnitude $\sim 1/(V\Sigma)$, i.e., $z=\lambda V\Sigma\sim{\cal
  O}(1)$.  From Eq.~(\ref{eq1.4}) it is intuitively clear that we will
only see an effect of the dynamical quarks on the microscopic spectral
correlations if the quark mass is also of this size.  For
convenience, we rescale the quark mass by the same factor as the
eigenvalues and define $\mu=mV\Sigma$.  For $\mu\gg1$, we should
simply obtain results which are identical to the quenched
approximation, since in this limit the fermion determinants do not
affect the small eigenvalues.  To observe results which are different
from those obtained in the quenched approximation, we require
$\mu\sim{\cal O}(1)$.

In the following, we summarize recent results obtained in
Ref.~\cite{Berb98c}.  Analytical RMT results for the microscopic
spectral density in the presence of massive dynamical quarks are
currently only known for the unitary (UE) and chiral unitary (chUE)
ensemble \cite{Damg98a}.  For the chSE, results are known in the
chiral limit.  We have \cite{Naga95,Berb98b,Ma98}
\begin{eqnarray}
  \rho_s(z)&\!\!\!=\!\!\!&z[J_\alpha^2(2z)-
  J_{\alpha+1}(2z)J_{\alpha-1}(2z)]\nonumber\\
  & & -\frac{1}{2}J_\alpha(2z)\int_0^{2z} dt J_\alpha(t)\:,
  \label{eq4.1}
\end{eqnarray}
where $\alpha=N_f+2|\nu|$ and $J$ denotes the Bessel function.  If
$\alpha$ is zero or an odd integer, RMT results for the distribution
of the smallest eigenvalue are known analytically
\cite{Forr93,Naga98}.  For even $\alpha\ne0$ (this is relevant in our
case), they can be obtained as an expansion in zonal polynomials
\cite{Kane93,Berb98c}.  For $\mu\ne0$, the RMT results for $\rho_s(z)$
and $P(\lambda_{\rm min})$ can be obtained numerically by constructing
skew-orthogonal polynomials which obey orthogonality relations
determined by a weight function involving the fermion determinants
\cite{Naga95,Berb98c}.  To avoid cancellation problems, a
multi-precision package was used \cite{Bail94}.

\begin{figure}[t]
  \centerline{\epsfig{figure=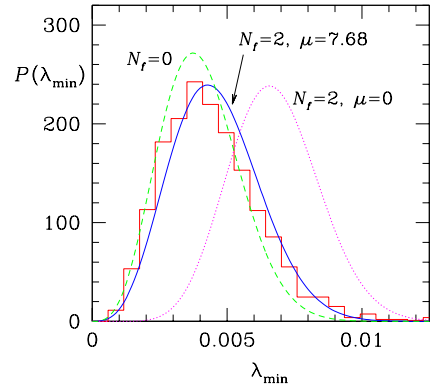,width=60mm}}
  \centerline{\epsfig{figure=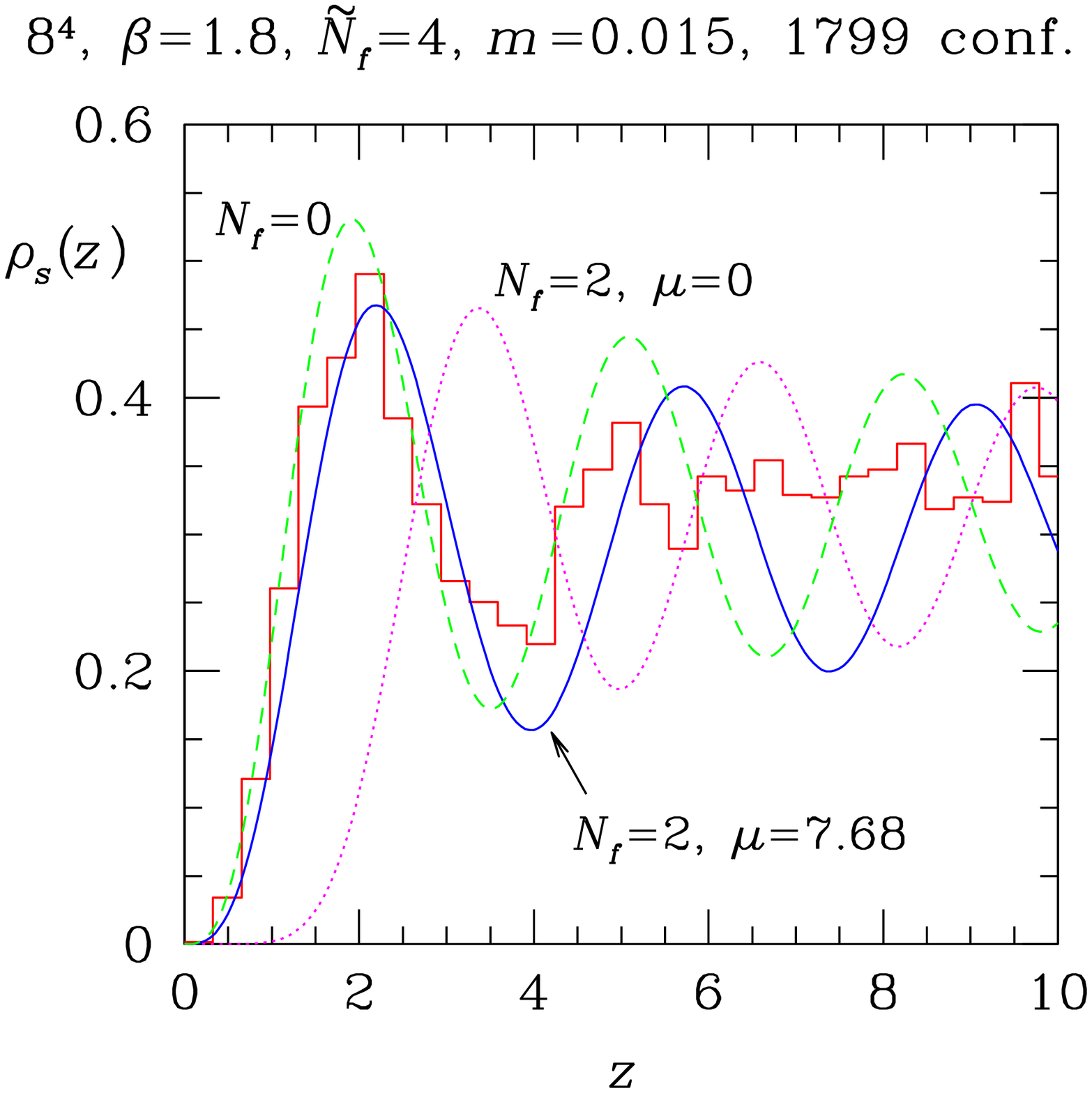,width=60mm}}
  \vspace*{-8mm}
  \caption{Distribution of the smallest eigenvalue (top) and
    microscopic spectral density (bottom) for the simulation
    parameters indicated above the figures.  The histograms represent
    the lattice data.  The solid curves are the RMT results computed
    with the appropriate value of the rescaled quark mass $\mu$.  The
    dashed and the dotted curves are the RMT results computed in the
    quenched approximation and in the chiral limit, respectively.}
  \label{fig2a}
\end{figure}

\begin{figure}[t]
  \centerline{\epsfig{figure=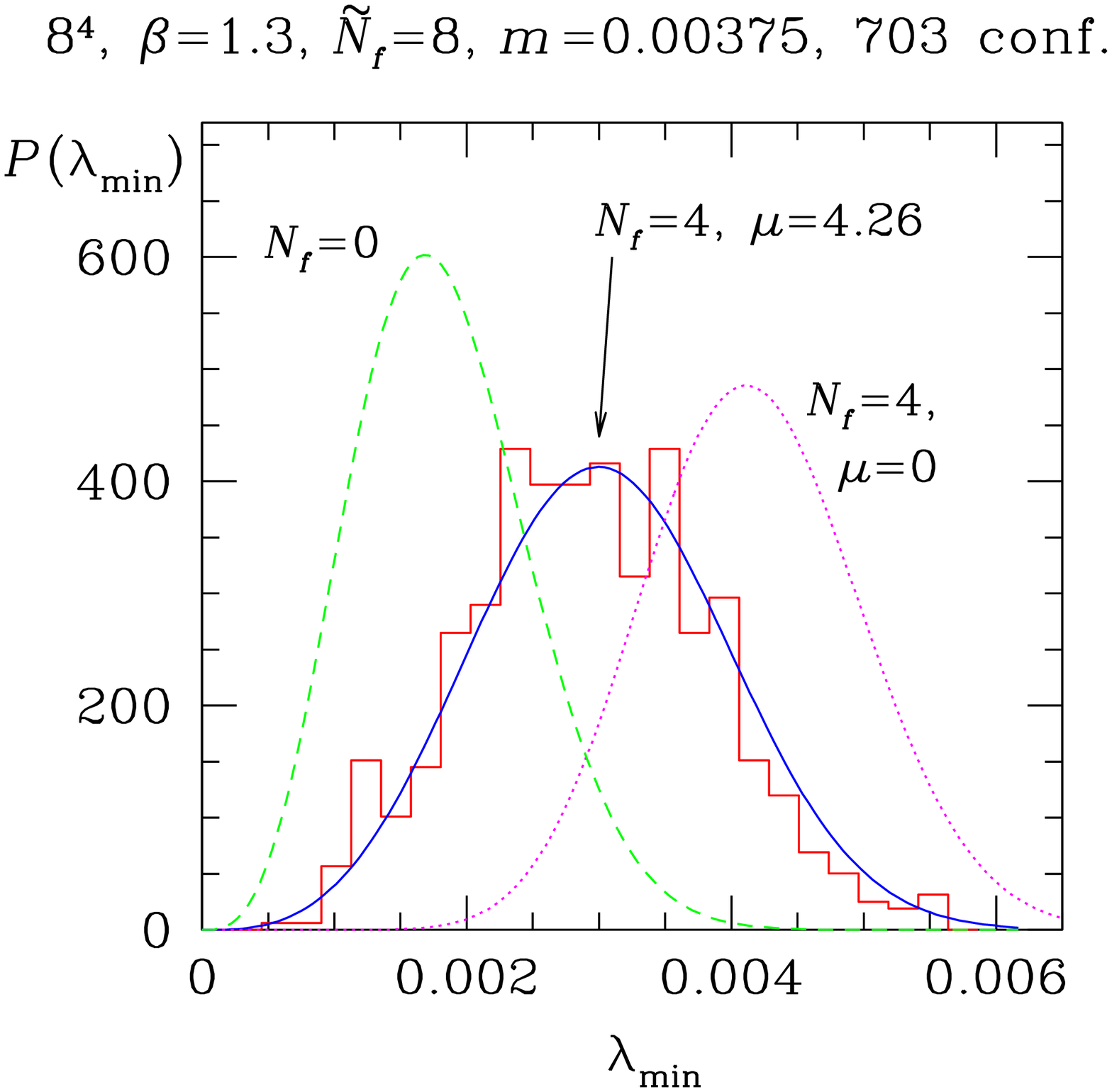,width=60mm}}
  \centerline{\epsfig{figure=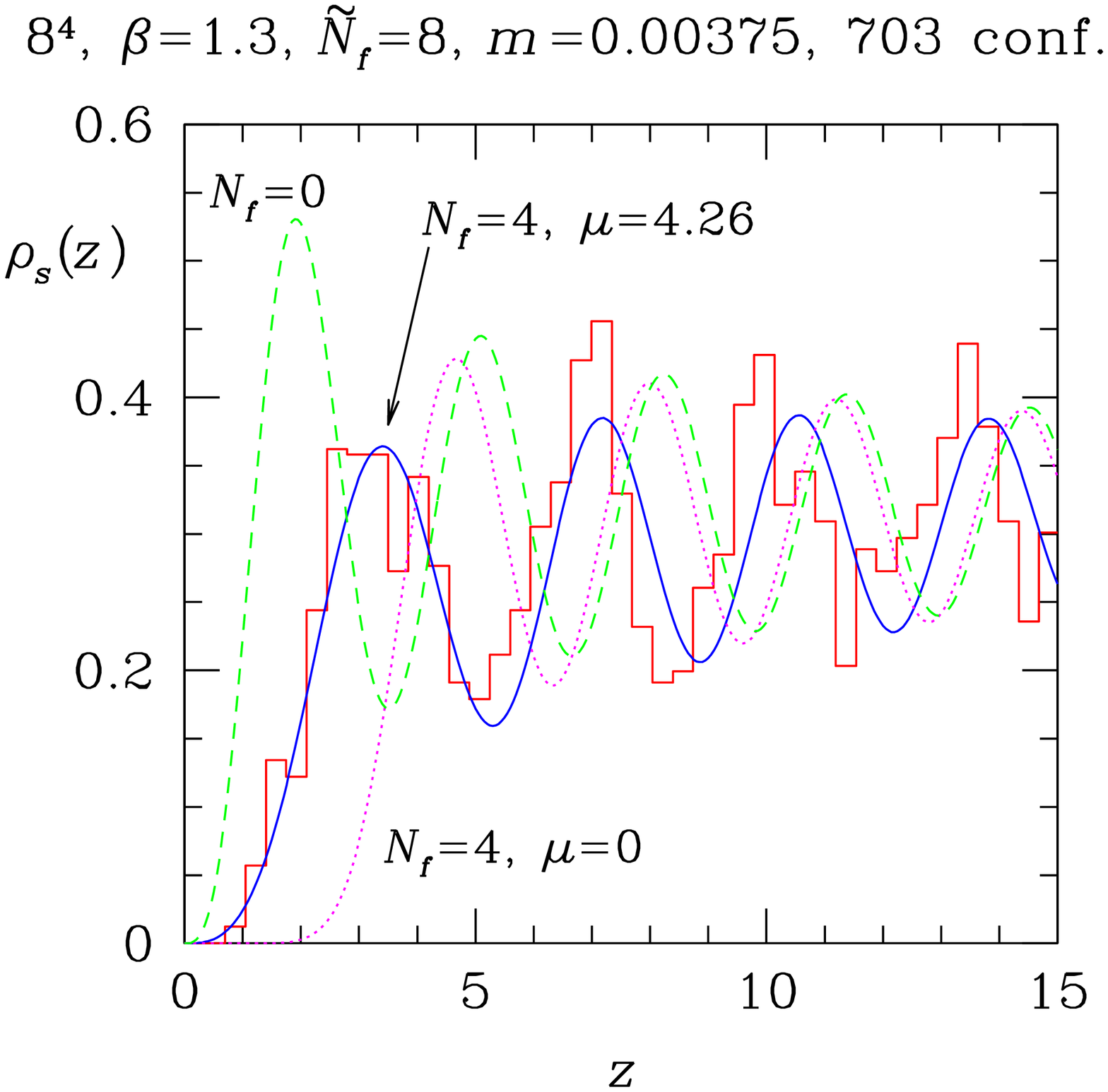,width=60mm}}
  \vspace*{-8mm}
  \caption{Same as Fig.~\ref{fig2a} but for different simulation
    parameters (indicated above the figures).}
  \label{fig2b}
\end{figure}

Some of our numerical results are shown in Figs.~\ref{fig2a} and
\ref{fig2b}.  In this first exploratory study, we chose values of
$\beta$ in the strong-coupling region since there one does not need
very large lattices to obtain agreement with RMT, see the end of
Sec.~\ref{quen} and Sec.~\ref{susc}.  The RMT results were computed
with $\nu=0$ for the reason discussed in Sec.~\ref{quen}.  Again,
there is only one parameter, $V\Sigma$, which sets the energy scale
and is determined by the data from the Banks-Casher formula.  We have
also plotted the RMT results for $\rho_s(z)$ and $P(\lambda_{\rm
  min})$ in the quenched approximation and in the chiral limit,
respectively.  The data should agree with these two curves in the
limits $\mu\gg1$ and $\mu=0$, respectively.  For $\mu\sim{\cal O}(1)$,
the data should lie somewhere in between these two limiting curves.
This is indeed observed in the figures.

An important remark is in order.  We are comparing lattice data
computed with $n$ copies of staggered flavors (corresponding to
$\tilde N_f=4n$ flavors in the continuum limit) with RMT results
computed for $N_f$ flavors, where $N_f=\tilde N_f/2$.  This is not an
ad-hoc prescription but justified by the combination of the following
two reasons.  First, the U(4)$\times$U(4) symmetry of the continuum
action (for $\tilde N_f=4$) is broken to U(1)$\times$U(1) at finite
lattice spacing $a$.  This would suggest to use $N_f={\tilde N}_f/4$
in the RMT results.  Second, all eigenvalues of $D$ are doubly
degenerate in SU(2).  However, in Eq.~(\ref{eq1.4}) the eigenvalues
are assumed to be non-degenerate \cite{Verb94a}.  This would suggest
to use $N_f=2{\tilde N}_f$ in the RMT results.  Combining the two
factors, we conclude that we have to use $N_f=\frac14\cdot2\cdot\tilde
N_f=\tilde N_f/2$ in the RMT results.

Note that in the regime $\mu\sim{\cal O}(1)$, the quark mass is very
small, and lattice simulations are very demanding.  Testing the RMT
predictions required a substantial numerical effort.  However, the
point is that, once the validity of the RMT description is
established, the availability of analytical results for very small
quark mass should facilitate extrapolations to the chiral limit which
are otherwise difficult to perform on the lattice.

\section{Crossover to non-universal behavior}
\label{susc}

As indicated above, the domain of validity of the random matrix
description is finite.  After all, QCD is a complicated theory and not
just a random matrix model.  If one wants to make practical use of the
RMT results, one should know quantitatively in which energy range they
apply.  In condensed matter physics, the energy up to which the RMT
description is valid is called the Thouless energy, $E_c$.
Essentially, $E_c$ is the inverse of the diffusion time of an electron
through a disordered mesoscopic sample.  It behaves like $E_c\sim
L^{-2}$, where $L$ is the length of the sample.  Recently, some
progress was made in identifying the equivalent of the Thouless energy
in QCD \cite{Ster98,Jani98,Osbo98,Berb98d,Guhr98}.  Again, we shall
concentrate on the low-lying eigenvalues.

The starting point is the condition for the applicability of the
Leutwyler-Smilga effective theory,
\begin{equation}
  \label{eq5.1}
  L<\frac{1}{m_\pi}\:,
\end{equation}
which also sets the domain of validity of the RMT description.  From
the Gell-Mann--Oakes--Renner relation, we have
\begin{equation}
  \label{eq5.2}
  m_\pi^2 f_\pi^2=2m\Sigma\:.
\end{equation}
Here, $m$ is to be regarded as a valence quark mass which sets the
energy scale below which RMT applies.  Denoting the QCD equivalent of
the Thouless energy by $\lambda_{\rm RMT}$, one obtains
\cite{Jani98,Osbo98}
\begin{equation}
  \label{eq5.3}
  \lambda_{\rm RMT}\sim\frac{f_\pi^2}{\Sigma L^2}\:.
\end{equation}
To have a dimensionless quantity, it is useful to divide $\lambda_{\rm
  RMT}$ by the mean level spacing at $\lambda=0$,
$\Delta=1/\rho(0)=\pi/(V\Sigma)$.  This yields
\begin{equation}
  \label{eq5.4}
  \lambda_{\rm RMT}/\Delta\sim\frac{1}{\pi}f_\pi^2L^2\:.
\end{equation}
This is the prediction which we want to test.  It was already checked
in the instanton liquid model and confirmed qualitatively
\cite{Osbo98}.  Since we have a large number of complete spectra at
our disposal, we are in a position to perform a comprehensive
quantitative test against lattice data.  For simplicity, we only use
the quenched data.  In the following, we summarize results obtained in
Ref.~\cite{Berb98d}.

\begin{figure}
  \centerline{\epsfig{figure=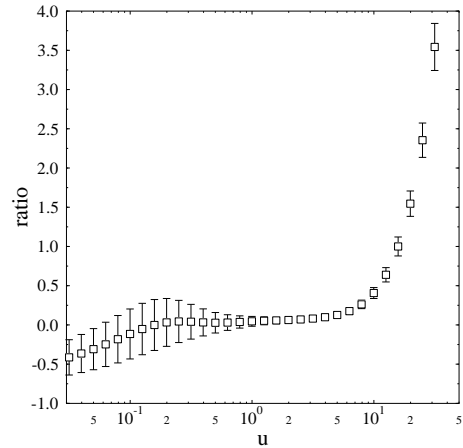,width=60mm}}
  \vspace*{-8mm}
  \caption{The ratio defined in Eq.~(\protect\ref{eq5.7}) versus $u$
    for the data obtained on a $10^4$ lattice using $\beta=2.0$.}
  \label{fig3}
\end{figure}

A convenient quantity to consider in this respect is the disconnected
chiral susceptibility.  It is defined in terms of the Dirac
eigenvalues by
\begin{eqnarray}
   \chi^{\rm disc}_{\rm lattice}&=&\frac{1}{N}\left\langle\sum_{k,l=1}^N
    \frac{1}{(i\lambda_k+m)(i\lambda_l+m)}\right\rangle\nonumber\\
  &&- \frac{1}{N}
  \left\langle\sum_{k=1}^N\frac{1}{i\lambda_k+m}\right\rangle^2\:,
  \label{eq5.5}
\end{eqnarray}
where $N=Va^{-4}$ is the number of lattice sites and $m$ is the
valence quark mass.  Going over to the microscopic limit, i.e.,
rescaling all energies by $V\Sigma$ and defining $u=mV\Sigma$, the
susceptibility can be expressed in terms of the microscopic spectral
one- and two-point functions.  Inserting the RMT results of the chSE
for these quantities and performing some tedious algebra, we obtain
the RMT prediction
\begin{eqnarray}
  \label{eq5.6}
  \chi^{\rm disc}_{\rm RMT}&\!\!\!\!\!=\!\!\!\!\!&
  4u^2 \int_0^1 ds\: s^2K_0(2su) \int_0^1 dt\: I_0(2stu)\nonumber\\
  &&\quad\times\Bigl\{s(1-t^2)-8stI_0(2stu)K_0(2su)\nonumber\\
  &&\qquad\;+4K_0(2u)\left[I_0(2su)+tI_0(2stu)\right]\Bigr\}\nonumber\\
  &&-4u^2K_0^2(2u) \left[ \int_0^1 ds \:I_0(2su)\right]^2,
\end{eqnarray}
where $I$ and $K$ denote modified Bessel functions.  Note that in
going from Eq.~(\ref{eq5.5}) to Eq.~(\ref{eq5.6}), $\chi^{\rm disc}$
has been rescaled by $1/(N\Sigma^2)$ to eliminate the dependence on
$\beta$.  The RMT result should agree with the lattice result up to
some limiting value $u_{\rm RMT}$.  To identify this value, it is
useful to define the ratio
\begin{equation}
  \label{eq5.7}
  {\rm ratio}=\frac{\chi^{\rm disc}_{\rm lattice}
    -\chi^{\rm disc}_{\rm RMT}}{\chi^{\rm disc}_{\rm RMT}}\:.
\end{equation}
Deviations of this ratio from zero indicate the breakdown of the RMT
description.  The ratio is plotted in Fig.~\ref{fig3} for one
particular data set.
The deviations of the ratio from zero for very small values of $u$ are
artifacts of finite statistics which are explained in
Ref.~\cite{Berb98d}.  We are interested in the deviations from zero
which set in at $u\approx5\sim10$.  Let us denote this value of $u$ by
$u_{\rm RMT}$.  It is related to $\lambda_{\rm RMT}$ by $u_{\rm
  RMT}=\lambda_{\rm RMT}V\Sigma=\pi\lambda_{\rm RMT}/\Delta$.

\begin{figure}
  \centerline{\epsfig{figure=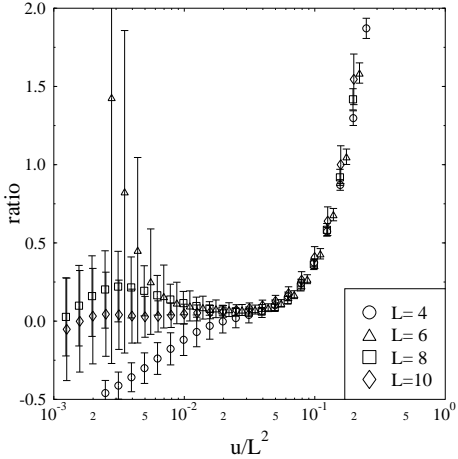,width=60mm}}
  \vspace*{-8mm}
  \caption{The ratio of Eq.~(\protect\ref{eq5.7}) versus $u/L^2$ for
    constant $\beta=2.0$ and four different lattice sizes $V=L^4$.}
  \label{fig4}
\end{figure}

According to Eq.~(\ref{eq5.4}), $u_{\rm RMT}$ should scale with $L^2$.
This prediction is tested in Fig.~\ref{fig4} where we have plotted
data for four different lattice sizes at constant $\beta$ versus
$u/L^2$.
All data fall on the same curve, except for the deviations at small
$u$ which are due to the finiteness of our statistical ensembles.
This confirms the prediction that $u_{\rm RMT}$ scales with $L^2$.

\begin{figure}[-t]
  \centerline{\epsfig{figure=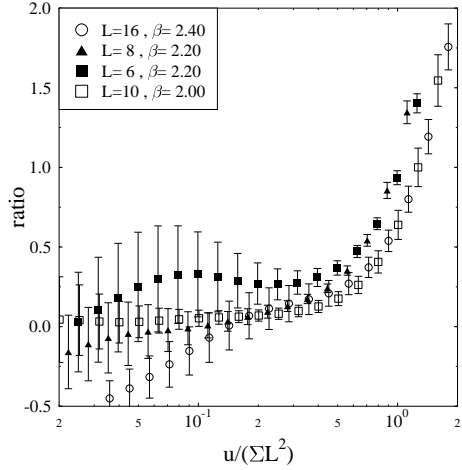,width=60mm}}
  \vspace*{-8mm}
  \caption{The ratio of Eq.~(\protect\ref{eq5.7}) versus $u/(\Sigma
    L^2)$ for four different parameter sets.}
  \label{fig5}
\end{figure}

It remains to test the predicted scaling with $f_\pi^2$.  In order to
determine $f_\pi$, we use a result of Ref.~\cite{Bill85} where it was
found that $f_\pi^2\approx3.4\Sigma$ in lattice units in the range of
$\beta$ we consider.  This suggests to plot the ratio of
Eq.~(\ref{eq5.7}) versus $u/(\Sigma L^2)$ which we have done in
Fig.~\ref{fig5}.
Again, the data fall approximately on the same curve, in particular
the data for $\beta=2.0$ and $\beta=2.4$.  (One can only expect an
agreement on the level to which the relation between $f_\pi^2$ and
$\Sigma$ is known.)  Putting in the numbers, we obtain
\begin{equation}
  \label{eq5.8}
  \lambda_{\rm RMT}/\Delta\approx0.3f_\pi^2L^2
\end{equation}
which is in good agreement with the prediction of Eq.~(\ref{eq5.4}).

Another interesting quantity is the number variance in an interval
$I=[0,S]$, defined by $\Sigma^2(I)=\langle(N(I)-\langle N(I)\rangle)^2
\rangle$. (This $\Sigma^2$ should not be confused with the absolute
value of the chiral condensate.)  Here, $N(I)$ is the number of
eigenvalues in $I$, and $\langle\cdots\rangle$ denotes an ensemble
average.  The number variance can be computed in RMT, and the lattice
data agree with the RMT result up to some limiting value of $S$ which
we will denote by $S_{\rm RMT}$.  Considering a number of different
lattice sizes and $\beta$-values we found that $S_{\rm RMT}$ scales
with $f_\pi^2$ and $L^2$ as expected.  Quantitatively,
\begin{equation}
  \label{eq5.9}
  S_{\rm RMT}\approx(0.3\sim0.7)f_\pi^2L^2
\end{equation}
which is consistent with $u_{\rm RMT}$.

We conclude this section with the statement that the domain of
validity of the RMT description of the low-lying Dirac eigenvalues in
QCD is now known quantitatively.  The implications of this result will
be discussed in the next section.

\section{Summary and outlook}
\label{summary}

We now have solid numerical evidence that the distribution of the
low-lying Dirac eigenvalues is universal and described by RMT results,
both in the quenched approximation and with dynamical fermions.
Furthermore, we have a quantitative criterion for the domain of
validity of the random matrix description which was confirmed by
lattice simulations.  How can we make practical use of this knowledge?

One point concerns extrapolations to limits that are difficult to take
on the lattice.  The first example is the thermodynamic limit.  Since
we are dealing with finite-volume effects, we can make analytical
statements about how this limit is approached.  This was already
demonstrated in Ref.~\cite{Berb98b}.  The second example is the chiral
limit.  We have RMT results for the microscopic spectral quantities in
the presence of dynamical quarks with arbitrarily small masses.  For
quantities that are sensitive to the small eigenvalues, these results
should provide guidance for extrapolations to the chiral limit.
Third, we have the continuum limit.  It is less clear in what way the
random matrix approach might help in approaching this limit.  One
observation is that as $a\to0$, there should be a transition in the
effective number of flavors used in the RMT results with dynamical
fermions, from $\tilde N_f/2$ to $2\tilde N_f$.  Presumably, $a$ has
to be very small to see such a transition.

As far as topology is concerned, RMT can determine the microscopic
spectral quantities only for fixed topological charge $\nu$.  The
overall result for some quantity will be a weighted average over all
topological sectors, and RMT cannot predict the weights.  So far, all
data were consistent with $\nu=0$ in the RMT results since the latter
are only sensitive to exact zero modes of the Dirac operator.  It
would be very interesting to identify the would-be zero modes.  For
this, one needs the eigenvectors corresponding to the small
eigenvalues.  We are currently trying to compute these as well.

Last but not least, we feel that the analytical knowledge of the
distribution of the smallest eigenvalues may have an impact on fermion
algorithms.  This is suggested by the fact that the magnitude of the
small eigenvalues determines the performance of the fermion algorithm.
While RMT cannot say anything about individual configurations, it
predicts the statistical distribution of the small eigenvalues.  This
knowledge may be useful to construct more effective algorithms.

\bigskip\noindent{\bf Acknowledgments.} This work was supported in
part by DFG grants Me 567/5-3 and We 655/15-1.  The numerical
simulations were done on a CRAY T3E-900 at the HLRS Stuttgart.  We
thank the MPI f\"ur Kernphysik, Heidelberg, for hospitality and
support.

\end{document}